\documentclass[a4paper]{article}

\author{G. Gubbiotti\footnote{e-mail:
gubbiotti@mat.uniroma3.it},$\quad$  C. Scimiterna\footnote{e-mail:
scimiterna@fis.uniroma3.it}, $\quad$ D. Levi\footnote{e-mail:
decio.levi@roma3.infn.it}
\\
 Dipartimento di Matematica e Fisica, Universita' degli Studi Roma Tre,\\ e Sezione INFN di Roma Tre,\\ Via della Vasca Navale 84, 00146 Roma (Italy)
 }
 \title{A non autonomous generalization of the $Q_{\text{V}}$ equation}
\usepackage{rotating,color}

\usepackage{cite, amssymb}
\usepackage{braket}
\usepackage{amsmath}
\usepackage{booktabs}
\usepackage{amscd}
\usepackage[Q=yes]{examplep}
\usepackage{multirow}
\usepackage{rotating}
\usepackage{array}
\usepackage{longtable}
\usepackage{pdflscape}
\usepackage{array}
\newcolumntype{C}{>{\displaystyle} c <{}}

\newcommand{\Hvier}{$H^{4}$}
\newcommand{\Hsechs}{$H^{6}$}

\newcommand{\Z}{\mathbb{Z}}

\def\bea{\begin{eqnarray}}
\def\eea{\end{eqnarray}}

\newcommand{\diff}[2]{\frac{\mathrm{d} #1}{\mathrm{d} #2}}
\newcommand\scalemath[2]{\scalebox{#1}{\mbox{\ensuremath{\displaystyle #2}}}}

\newcommand{\Fp}[1]{F^{(+)}_{#1}}
\newcommand{\Fm}[1]{F^{(-)}_{#1}}

\newcommand{\QV}{Q_{\text{V}}}

\newcounter{rmk}
\renewcommand{\thermk}{\arabic{rmk}}
{\refstepcounter{rmk}\vspace{6pt}\noindent\ignorespaces\textbf{Remark
\thermk:}}{\vspace{6pt}\par}

\renewcommand{\epsilon}{\varepsilon}

\allowdisplaybreaks

\begin{document}

\maketitle

\begin{abstract}
    In this paper we introduce a non autonomous generalization of the
    $\QV$ equation introduced by Viallet. All  the equations of
     Boll's classification  appear in it for   special choices of the parameters. Using the algebraic
    entropy test we infer that the equation should be integrable and with the
    aid of a formula introduced by Xenitidis we find  its three point generalized
    symmetries.
\end{abstract}

\section{Introduction}

In a previous paper \cite{GubScimLev2015b} we calculated the three point
generalized symmetries of equations belonging to Boll classification \cite{Boll2011,Boll2012a,Boll2012b} of quad graph equations compatible around the cube, i.e.  the
trapezoidal \Hvier~equations and the \Hsechs~equations.  The symmetres of the rhombic \Hvier~equations has been considered  in \cite{XenitidisPapageorgiou2009}. In \cite{GubScimLev2015b} we  also noticed that all
the fluxes of such symmetries are related to some particular cases of the non 
autonomous YdKN equation \cite{LeviYamilov1997}\footnote{For the rhombic
    \Hvier~equations this was already known from \cite{XenitidisPapageorgiou2009}.}:
\begin{equation}
    \diff{u_{n}}{t} = \frac{A_{n}(u_{n})u_{n+1}u_{n-1}
    + B_{n}(u_{n})(u_{n+1}+u_{n-1}) + C_{n}(u_{n})}{u_{n+1}-u_{n-1}}.
    \label{eq:naydkn}
\end{equation}
Here the $n$-dependent coefficients are given by:
\begin{subequations}
\begin{align}
    A_{n}(u_n)&= a u_{n}^{2} + 2 b_{n} u_{n} + c_{n},
    \\
    B_{n}(u_n) &= b_{n+1} u_{n}^{2} + d u_{n} + e_{n+1},
    \\
    C_{n} (u_n)&= c_{n+1}u_{n}^{2} + 2 e_{n} u_{n} + f,
\end{align}
    \label{eq:akbkcknaydkn}%
\end{subequations}
where $b_{n}$, $c_{n}$ and $e_{n}$ are 2-periodic functions, i.e. 
\bea \label{aYdKN}
b_n=b_0+b_1 (-1)^n,\; c_n=c_0+c_1 (-1)^n, \; e_n=e_0+e_1 (-1)^n. 
\eea 
\cite{GubScimLev2015b}  naturally extended the results contained  in \cite{LeviPetreraScimiterna2008} where the three points generalized symmetries
of the ABS class of lattice equations \cite{ABS2003}, which are at the  base
of  Boll's classification, are shown to be  sub--cases of  the autonomous YdKN equation
\cite{Yamilov1983}. We recall
that the autonomous YdKN equation is obtained from \eqref{eq:naydkn} just taking
$b_{k}$, $c_{k}$ and $e_{k}$ as pure constants, i.e.:
\begin{equation}
      \diff{u_{n}}{t} = \frac{A(u_{n})u_{n+1}u_{n-1}
    + B(u_{n})(u_{n+1}+u_{n-1}) + C(u_{n})}{u_{n+1}-u_{n-1}}.
    \label{eq:ydkn}
\end{equation}
where:
\begin{subequations}
\begin{align}
     A(u_n)&= a u_{n}^{2} + 2 b u_{n} + c,
    \\
    B(u_n) &= b u_{n}^{2} + d u_{n} + e,
    \\
    C (u_n)&= cu_{n}^{2} + 2 e u_{n} + f,
\end{align}
\label{eq:abcydkn}%
\end{subequations}
 In particular in \cite{GubScimLev2015b} we noted that all equations in the Boll classification are sub-cases of the general non autonomous YdKN (\ref{eq:naydkn}) corresponding to $a=b_{k}=0$.

In \cite{Xenitidis2009} it was shown that the 
$\QV$ equation, introduced in \cite{VialletQV}:
\begin{equation}
    \begin{aligned}
    \QV = & a_{1} u_{{n,m}}u_{{n+1,m}}u_{{n,m+1}}u_{{n+1,m+1}} 
    \\
    + &a_{2,0}\left(u_{{n,m}}u_{{n,m+1}}u_{{n+1,m+1}}+ u_{{n+1,m}}u_{{n,m+1}}u_{{n+1,m+1}}\right.
    \\
    &\phantom{a_{2}}\left.+u_{{n,m}}u_{{n+1,m}}u_{{n+1,m+1}}+u_{{n,m}}u_{{n+1,m}}u_{{n,m+1}}\right)
    \\
    + &{ a_{3,0}} \left(u_{{n,m}}u_{{n+1,m}}+ u_{{n,m+1}}u_{{n+1,m+1}}\right)
    \\
    + &a_{4,0}\left(u_{{n,m}}u_{{n+1,m+1}}+ u_{{n+1,m}}u_{{n,m+1}}\right)
    \\
    + &a_{5,0}\left(u_{{n+1,m}}u_{{n+1,m+1}}+u_{{n,m}}u_{{n,m+1}}\right)
    \\
    + &a_{6,0}\left( u_{{n,m}}+u_{{n+1,m}}+u_{{n,m+1}}+ u_{{n+1,m+1}}\right)
    \\
    +&a_{7}=0
    \end{aligned}
    \label{eq:qv}
\end{equation}
admits a symmetry in the direction $n$ 
\begin{equation}
\frac{du_{n,m}}{dt} = \frac {h_n}{u_{n+1,m} - u_{n-1,m}} 
    - \frac{1}{2} \partial_{u_{n+1,m}}h_n. 
    \label{eq:ttxsymmn}
\end{equation}
where:
\begin{equation}
    \label{Vn}
    \begin{aligned}
    h_n(u_{n,m}, u_{n+1,m}) &= Q_{V}  \partial_{u_{n,m+1}} \partial_{u_{n+1,m+1}} Q_{V}
    \\
    &-\left(  \partial_{u_{n,m+1}} Q_{V} \right) \left(   \partial_{u_{n+1,m+1}} Q_{V}  \right)    
    \end{aligned}
\end{equation}
and a symmetry in the direction $m$ 
\begin{equation}
\frac{du_{n,m}}{dt} = \frac {h_m}{u_{n,m+1} - u_{n,m-1}} 
    - \frac{1}{2} \partial_{u_{n,m+1}}h_m. 
    \label{eq:ttxsymmm}
\end{equation}
where:
\begin{equation}
    \label{Vm}
    \begin{aligned}
    h_m(u_{n,m}, u_{n,m+1}) &= Q_{V}  \partial_{u_{n+1,m}} \partial_{u_{n+1,m+1}} Q_{V}
    \\
    &-\left(  \partial_{u_{n+1,m}} Q_{V} \right) \left(   \partial_{u_{n+1,m+1}} Q_{V}  \right) 
    \end{aligned}
\end{equation}
of the same form as the YdKN (\ref{eq:ydkn}). 

The connection formulae between the coefficient of $Q_V$ and the $n$ direction YdKN \eqref{eq:ydkn} are:
\bea \nonumber
a= a_{3,0} a_1-a_{2,0}^2, \qquad b=\frac 1 2 [ a_{2,0} (a_{3,0}-a_{5,0}-a_{4,0})+a_{6,0} a_1], \\ \label{QV-YdKNn} c=  a_{2,0} a_{6,0}- a_{4,0} a_{5,0}, \qquad d=\frac 1 2 [ a_{3,0}^2-a_{4,0}^2-a_{5,0}^2 +a_1 a_7], \\ \nonumber
e=\frac 1 2 [ a_{6,0} (a_{3,0}-a_{4,0}-a_{5,0})+a_{2,0} a_7], \qquad f= a_{3,0} a_7-a_{6,0}^2.
\eea
and show that $\left(a,b\right)$ may be different from $\left(0,0\right)$. The connection formulae between the coefficient of $Q_V$ and the $m$ direction YdKN \eqref{eq:ydkn} are:
\bea \nonumber
a= a_{5,0} a_1-a_{2,0}^2, \qquad b=\frac 1 2 [ a_{2,0} (a_{5,0}-a_{3,0}-a_{4,0})+a_{6,0} a_1], \\ \label{QV-YdKNm} c=  a_{2,0} a_{6,0}- a_{4,0} a_{3,0}, \qquad d=\frac 1 2 [ a_{5,0}^2-a_{4,0}^2-a_{3,0}^2 +a_1 a_7], \\ \nonumber
e=\frac 1 2 [ a_{6,0} (a_{5,0}-a_{4,0}-a_{3,0})+a_{2,0} a_7], \qquad f= a_{5,0} a_7-a_{6,0}^2.
\eea
 So  the three point generalized symmetries
of the $\QV$ belong to the class of the general YdKN equation \eqref{eq:ydkn}.

From the results obtained in \cite{GubScimLev2015b} we 
were led to conjecture the existence of a non autonomous generalization 
of the $\QV$ equation and suggested some possible ways to obtain such a 
generalization. In this paper we follow them to   generalize 
the $\QV$ equation. It is known  \cite{VialletQV} that $\QV$ is the most general multi-linear 
equation on a quad graph possessing Klein discrete symmetries, i.e. such that:
\begin{equation}
    \begin{aligned}
        Q\left(u_{n+1,m},u_{n,m},u_{n+1,m+1},u_{n,m+1}\right)=
        \tau Q\left(u_{n,m},u_{n+1,m},u_{n,m+1},u_{n+1,m+1}\right),
        \\
        Q\left(u_{n,m+1},u_{n+1,m+1},u_{n,m},u_{n,m+1}\right)=
        \tau^{\prime}Q\left(u_{n,m},u_{n+1,m},u_{n,m+1},u_{n+1,m+1}\right),        
    \end{aligned}
    \label{eq:Klein}%
\end{equation}
where $\tau,\tau^{\prime}=\pm 1$. A  way to  extend 
the Klein discrete symmetry is to consider a 
multi-linear function $Q$ with  two-periodic coefficients in $n$ and $m$ such that the following equations holds:
\begin{equation}
    \begin{aligned}
        &Q\left(u_{n+1,m},u_{n,m},u_{n+1,m+1},u_{n,m+1};(-1)^n,(-1)^m\right)=
        \\
        &\phantom{=}\tau
        Q\left(u_{n,m},u_{n+1,m},u_{n,m+1},u_{n+1,m+1};-(-1)^n,(-1)^m\right),
        \\
        &Q\left(u_{n,m+1},u_{n+1,m+1},u_{n,m},u_{n+1,m};(-1)^n,(-1)^m\right)=
        \\
        &\phantom{=}\tau'Q\left(u_{n,m},u_{n+1,m},u_{n,m+1},u_{n+1,m+1};(-1)^n,-(-1)^m\right).
    \end{aligned}
    \label{KL}
\end{equation}
If a non autonomous system $Q$ satisfies the discrete symmetries (\ref{KL})   we will say that $Q$ admits a \emph{non autonomous discrete
Klein symmetry}. The name follows from the fact  that if  $Q$ is autonomous then the
discrete symmetry \eqref{KL} reduces to the Klein one \eqref{eq:Klein}.
Furthermore  
all the equation belonging to the Boll's classification satisfy this symmetry conditions
\eqref{KL} when $\tau=\tau^{\prime}=1$.

In Section 2 we will show that up to multiplication by a function
and  re-definition of the parameters there is only one quad graph equation
 possessing the non autonomous Klein symmetry.
We will prove that this equation, which we will call the \emph{non autonomous
$\QV$ equation} has as sub--cases all the equations of  Boll classification and 
it satisfies the Algebraic Entropy integrability test. 
Furthermore we will show that  (\ref{eq:ttxsymmn}, \ref{Vn})   will provide an $n$ directional  and (\ref{eq:ttxsymmm}, \ref{Vm})   an $m$ directional symmetry for the non autonomous
$\QV$ equation and that such symmetries belong to the class of
the  non autonomous YdKN equation (\ref{eq:naydkn}, \ref{eq:akbkcknaydkn}). 
We will provide then the appropriate connection formul\ae. In Section 3 we will present some concluding remarks.

\section{The non autonomous $\QV$ equation}

Let us consider the most general multi-linear equation in the lattice
variables with two-periodic coefficients:
\begin{equation}
    \begin{aligned}
        &p_1 u_{{n,m}}u_{{n+1,m}}u_{{n,m+1}}u_{{n+1,m+1}}
    \\
    +&p_2 u_{{n,m}}u_{{n,m+1}}u_{{n+1,m+1}}+ 
    p_3 u_{{n+1,m}}u_{{n,m+1}}u_{{n+1,m+1}}
    \\
    +&p_4 u_{{n,m}}u_{{n+1,m}}u_{{n+1,m+1}}+
    p_5 u_{{n,m}}u_{{n+1,m}}u_{{n,m+1}}
    \\
    +&p_6 u_{{n,m}}u_{{n+1,m}}+ 
    p_7 u_{{n,m+1}}u_{{n+1,m+1}}+ 
    p_8 u_{{n,m}}u_{{n+1,m+1}}
    \\
    +&p_9 u_{{n+1,m}}u_{{n,m+1}}+
    p_{10} u_{{n+1,m}}u_{{n+1,m+1}}+ 
    p_{11} u_{{n,m}}u_{{n,m+1}}
    \\
    +&p_{12} u_{{n,m}}+ 
    p_{13}u_{{n+1,m}}+ 
    p_{14}u_{{n,m+1}}+
    p_{15} u_{{n+1,m+1}}
    +p_{16}=0
    \end{aligned}
    \label{eq:multilingen}%
\end{equation}
i.e. the $p_{i}$ coefficients have the following expression:
\begin{equation}
    p_{i} = p_{i,0} + p_{i,1}(-1)^{n} + p_{i,2} (-1)^{m} + p_{i,3}(-1)^{n+m}
    ,\quad i=1,\dots,16.
    \label{eq:pi}
\end{equation}
If we impose the non autonomous Klein symmetry condition \eqref{KL} with $\tau=\tau'=1$ the 64 coefficients of (\ref{eq:multilingen}) turn out to be related among themselves and we can choose among them 16 independent coefficients. 
In term of the 16 independent coefficients (\ref{eq:multilingen})
reads:
\begin{equation}
    \begin{aligned}
    &a_{1} u_{{n,m}}u_{{n+1,m}}u_{{n,m+1}}u_{{n+1,m+1}} 
    \\
    + &\left[ { a_{2,0}}- \left( -1 \right) ^{n}{ a_{2,1}}
    - \left( -1 \right) ^{m}{a_{2,2}}+ \left( -1 \right) ^{n+m}{ a_{2,3}}\right]
    u_{{n,m}}u_{{n,m+1}}u_{{n+1,m+1}}
    \\
    + &\left[ { a_{2,0}}+\left( -1 \right) ^{n}{ a_{2,1}}- \left( -1 \right) ^{m}{ a_{2,2}}-
    \left( -1 \right) ^{n+m}{ a_{2,3}} \right]
    u_{{n+1,m}}u_{{n,m+1}}u_{{n+1,m+1}}
    \\
    + &\left[ { a_{2,0}}+ \left( -1 \right) ^{n}{ a_{2,1}}
    + \left( -1 \right) ^{m}{ a_{2,2}}+ \left( -1 \right) ^{n+m}{ a_{2,3}} \right]
    u_{{n,m}}u_{{n+1,m}}u_{{n+1,m+1}}
    \\
    + &\left[ { a_{2,0}}- \left( -1 \right) ^{n}{a_{2,1}}+ \left( -1 \right) ^{m}{ a_{2,2}}
    - \left( -1 \right) ^{n+m}{ a_{2,3}} \right] 
    u_{{n,m}}u_{{n+1,m}}u_{{n,m+1}}
    \\
    + &\left[ { a_{3,0}}- \left( -1 \right) ^{m}{ a_{3,2}} \right] u_{{n,m}}u_{{n+1,m}}
    \\ 
    + &\left[ { a_{3,0}}+ \left( -1 \right) ^{m}{ a_{3,2}} \right] u_{{n,m+1}}u_{{n+1,m+1}}
    \\
    + &\left[ { a_{4,0}}- \left( -1\right) ^{n+m}{ a_{4,3}} \right] u_{{n,m}}u_{{n+1,m+1}}
    \\
    + &\left[ { a_{4,0}}+ \left( -1 \right) ^{n+m}{ a_{4,3}} \right] u_{{n+1,m}}u_{{n,m+1}}
    \\
    + &\left[ { a_{5,0}}- \left( -1 \right) ^{n}{ a_{5,1}} \right] u_{{n+1,m}}u_{{n+1,m+1}}
    \\
    + &\left[ { a_{5,0}}+ \left( -1 \right) ^{n}{ a_{5,1}} \right] u_{{n,m}}u_{{n,m+1}}
    \\
    + &\left[ { a_{6,0}}+ \left( -1 \right) ^{n}{ a_{6,1}}- 
    \left( -1 \right) ^{m}{ a_{6,2}}- \left( -1 \right) ^{n+m}{ a_{6,3}} \right] u_{{n,m}}
    \\
    + &\left[ { a_{6,0}}- \left( -1 \right) ^{n}{ a_{6,1}}
    - \left( -1 \right) ^{m}{ a_{6,2}}+ \left( -1 \right) ^{n+m}{ a_{6,3}} \right] u_{{n+1,m}}
    \\
    + &\left[ { a_{6,0}}+ \left( -1 \right) ^{n}{ a_{6,1}}
    + \left( -1 \right) ^{m}{ a_{6,2}}+ \left( -1 \right) ^{n+m}{ a_{6,3}} \right] u_{{n,m+1}}
    \\
    + &\left[ { a_{6,0}}- \left( -1 \right) ^{n}{ a_{6,1}}
    + \left( -1 \right) ^{m}{ a_{6,2}}- \left( -1 \right) ^{n+m}{ a_{6,3}}\right] u_{{n+1,m+1}}
    \\
    +&a_{7}=0
    \end{aligned}
    \label{eq:naqv}
\end{equation}
Upon the substitution $a_{2,1}=a_{2,2}=a_{2,3}=a_{3,2}=a_{4,3}=a_{5,1}=a_{6,1}=a_{6,2}=a_{6,3}=0$ 
(\ref{eq:naqv})  reduces to the $\QV$ equation (\ref{eq:qv}).  Therefore we will call  \eqref{eq:naqv}
\emph{the non autonomous $\QV$ equation}.

If we impose the non autonomous Klein symmetry condition \eqref{KL}
with the choice $\tau=1$ and $\tau'=-1$ we will get an expression
which can be reduced to \eqref{eq:naqv} by multiplying by $(-1)^{n}$ and 
redefining the coefficients. In an analogous manner the two remaining 
cases $\tau=-1$, $\tau'=1$ and  $\tau=\tau'=-1$ can be identified with the case 
$\tau=\tau'=1$ multiplying by $(-1)^{n}$ and $(-1)^{n+m}$ respectively and redefining the coefficients.
Therefore the only equation belonging to the class of the lattice equation
possessing the non autonomous Klein symmetries is just the non autonomous
$\QV$ equation \eqref{eq:naqv}.

We note that the non autonomous $\QV$ equation contains
as particular cases the rhombic \Hvier~equations, the trapezoidal \Hvier~
equations and the \Hsechs~equations. The explicit identification 
of the coefficients of such equations is given in Table \ref{tab:identification}.
The reader can refer to Appendix \ref{app:equations} for the explicit
expressions of these equations or to \cite{GubScimLev2015} for a complete
derivation following the prescription of \cite{Boll2012b}.

In order to establish if  \eqref{eq:naqv} is integrable
we use the Algebraic Entropy integrability test 
\cite{Viallet2006,HietarintaViallet2007,Tremblay2001},
using the program \texttt{ae2d.py} \cite{GubHay,gubbiotti_thesis}.
Applying it to the non autonomous $\QV$ equation we find  the following
degree of growth in all directions:
\begin{equation}
    1,3,7,13,21,31,43,57,73,91,111,133\dots,
    \label{eq:growthnaqv}
\end{equation}
which is the same as for the autonomous $\QV$ equation \cite{VialletQV}.
The generating function for the sequence \eqref{eq:growthnaqv}
is:
\begin{equation}
    g(z) = \frac{1+z^{2}}{(1-z)^{3}},
    \label{eq:genfuncnaqv}
\end{equation}
which implies  that we have
the following quadratic fit for the growth:
\begin{equation}
    d_{k} = k(k+1)+1,
    \label{eq:dknaqv}
\end{equation}
and thus the Algebraic Entropy is zero.
This is a strong indication of the integrability
of the non autonomous $\QV$ equation \eqref{eq:naqv}.

Using (\ref{eq:ttxsymmn}, \ref{Vn}) or (\ref{eq:ttxsymmm}, \ref{Vm}) with $\QV$ substituted by its non autonomous version we get a version of the  non autonomous YdKN (\ref{eq:naydkn}) however the proof that this is effectively a symmetry of the non autonomous $\QV$ encounters serious computational difficulties. 

We can prove by a direct computation its validity for the following sub-cases:
\begin{itemize}
\item When Qv equation is non autonomous with respect to one direction only, either $n$ or $m$. All the trapezoidal $_t H^{4}$ equations belong to these two sub-classes;
\item For all the $H^{6}$ equations, which are non autonomous in both directions.
\end{itemize}
Its validity for the autonomous Qv and for all the rhombic $_r H^{4}$ equations was already showed respectively in \cite{Xenitidis2009} and \cite{XenitidisPapageorgiou2009}. However we cannot prove its validity for the general case (\ref{eq:naqv}).

Here  in the following we compute  the connection formulae for the general non autonomous case (\ref{eq:naqv}). For the $n$ directional symmetry we have:
\bea \label{c1}
a&=&a_1 a_{3,0}- a_{2,0}^2  + \\ \nonumber && \quad +a_{2,1}^2 -  a_{2,2}^2 + 
  a_{2,3}^2 - (-1)^m (2 a_{2,0} a_{2,2} - 2  a_{2,1} a_{2,3}  +   a_1 a_{3,2}), \\ \nonumber b_0&=&\frac 1 2 \{a_{2,0} (a_{3,0} -  a_{5,0} -  a_{4,0}) + 
 a_1 a_{6,0} + a_{2,2} a_{3,2}   - a_{2,3} a_{4,3} - a_{2,1} a_{5,1} - \\ \nonumber &&\; - (-1)^m[ a_{2,2} (a_{5,0}  +  a_{3,0}+ a_{4,0})+a_{2,3} a_{5,1}  + a_1 a_{6,2}  + a_{2,0} a_{3,2}  +\\ \nonumber && \quad + a_{2,1} a_{4,3}]\}, \\ \nonumber 
 b_1&=&\frac 1 2 \{ a_{2,1} (a_{3,0}  -  a_{4,0}+  a_{5,0}) + a_{2,3} a_{3,2}  - a_{2,2} a_{4,3}   + a_{2,0} a_{5,1}  - 
 a_1 a_{6,1} +\\ \nonumber && \; + (-1)^m [a_1 a_{6,3} -  a_{2,3} (a_{3,0} + a_{4,0}-  a_{5,0}) - a_{2,1} a_{3,2} -  a_{2,0} a_{4,3}+\\ \nonumber && \quad +  a_{2,2} a_{5,1}] \},\\ \nonumber 
 c_0&=&a_{2,0} a_{6,0} - a_{4,0} a_{5,0}- 
 a_{2,1} a_{6,1} - a_{2,3} a_{6,3}+ 
 a_{2,2} a_{6,2}+\\ \nonumber &&\; - (-1)^m [a_{2,2} a_{6,0} -  a_{4,3} a_{5,1}  - a_{2,3} a_{6,1} + a_{2,0} a_{6,2}  -  a_{2,1} a_{6,3} ], \\ \nonumber
 c_1&=& a_{4,0} a_{5,1}+ a_{2,1} a_{6,0}-  
 a_{2,0} a_{6,1}+a_{2,3} a_{6,2}-a_{2,2} a_{6,3}+\\ \nonumber && \; +(-1)^m[a_{2,2} a_{6,1}- a_{4,3} a_{5,0}   - a_{2,3} a_{6,0}  -   a_{2,1} a_{6,2}  
  + a_{2,0} a_{6,3}], \\ \nonumber 
  d&=&\frac 1 2 [a_{3,0}^2 - a_{4,0}^2- a_{5,0}^2+ a_1 a_7- a_{3,2}^2  + a_{4,3}^2  + a_{5,1}^2 -\\ \nonumber && \; - 4 (-1)^m (a_{2,2} a_{6,0} + a_{2,3} a_{6,1} + a_{2,0} a_{6,2} + a_{2,1} a_{6,3} ], \\ \nonumber 
  e_0&=&\frac 1 2 \{ a_{6,0} [a_{3,0}- a_{4,0}  - a_{5,0} ]+ a_{2,0} a_7+ 
 a_{5,1} a_{6,1}- 
 a_{3,2} a_{6,2}+ 
 a_{4,3} a_{6,3}+\\ \nonumber && \; + (-1)^m [a_{3,2} a_{6,0}  +  a_{4,3} a_{6,1} +  a_{5,1} a_{6,3} - a_{6,2} (a_{3,0}   + a_{4,0}  + a_{5,0} )    -  a_{2,2} a_7]\},\\ \nonumber 
 e_1&=&\frac 1 2 \{ a_{6,1}[a_{3,0}- a_{4,0}+ 
 a_{5,0} ] - a_{5,1} a_{6,0}+ a_{4,3} a_{6,2}- 
 a_{3,2} a_{6,3}- a_{2,1} a_7+\\ \nonumber && \;+(-1)^m[ a_{4,3} a_{6,0}   +  a_{3,2} a_{6,1}   -  a_{5,1} a_{6,2}   +   a_{6,3}(a_{5,0} -  a_{3,0}-  a_{4,0})  +  a_{2,3} a_7\}, \\ \nonumber 
 f&=& 
 a_{3,0} a_7-a_{6,0}^2- a_{6,2}^2+ a_{6,3}^2+ a_{6,1}^2-(-1)^m(  2  a_{6,0} a_{6,2}  - 2  a_{6,1} a_{6,3}   -  a_{3,2} a_7).
  \eea  
  The non autonomous $\QV$ is not symmetric in the exchange of $n$ and $m$ so its symmetries in the $m$ direction are different for their dependence on the coefficients and so are the connection formulae, however, for the sake of the reader as they are not essentially different from those presente above we do not write them here but present them in  Appendix B, cfr. (\ref{c2}).
\begin{sidewaystable}[hbt]
    \centering
    \[
        \scalemath{0.65}{
        \begin{array}{CCCCCCCCCCCC}
            \toprule
            \text{Eq.} & a_{3,0} & a_{3,2} & a_{4,0} & a_{4,3} & a_{5,0}
            & a_{5,1} & a_{6,0} & a_{6,1} & a_{6,2}& a_{6,3} & a_{7}
            \\
            \midrule
            _{r}H_{1}^{\varepsilon} & 1 & 0 & \frac{1}{2}\varepsilon(\alpha-\beta) 
            & \frac{1}{2}\varepsilon(\alpha-\beta) &
            -1 & 0 & 0 & 0 & 0 & 0 & \beta-\alpha
            \\
            _{r}H_{2}^{\varepsilon} & 1 & 0 & 2\epsilon \left( \beta-\alpha\right) &
            2\epsilon \left(\beta -\alpha\right) & -1 & 0 &
            - \left( \alpha-\beta \right)  \left( \epsilon\alpha+1+\epsilon\beta \right) &
            0 & 0 & \epsilon \left( \beta^{2} - \alpha^{2}\right) &
            - \left( \alpha-\beta \right)  \left( 2\epsilon{\alpha}^{2}+\alpha
            +2\epsilon{\beta}^{2}+\beta \right)
            \\
            _{r}H_{3}^{\varepsilon} & \alpha& 0 & 
            \frac{1}{2}{\frac{\epsilon \left( \beta^{2} - \alpha^{2}\right) }{\alpha\beta}}
            &\frac{1}{2}{\frac{\epsilon \left( \beta^{2} - \alpha^{2}\right) }{\alpha\beta}}
            &-\beta& 0& 0 & 0& 0& 0& \delta \left( \alpha^{2}-\beta^{2} \right)
            \\
            _{t}H^{\varepsilon}_{1} & -\frac{1}{2}{ \alpha_{2}}{\epsilon}^{2}& -\frac{1}{2}{ \alpha_{2}}{\epsilon}^{2}
            & -1& 0 & 1 & 0 & 0& 0& 0& 0& -{ \alpha_{2}}
            \\
            _{t}H^{\varepsilon}_{2} & \epsilon{ \alpha_{2}}& \epsilon{ \alpha_{2}} &
            -1 & 0& 1 & 0 & \frac{1}{2}\alpha_2 \left[2+\epsilon(2\alpha_2+\alpha_3)\right]&
            0 & \epsilon{ \alpha_{2}}{ \alpha_{3}}+\frac{1}{2}\epsilon{{ \alpha_{2}}}^{2}& 0 &
            \alpha_{2} \left[ \alpha_{2}+2\alpha_{3}+
            \varepsilon \left( \alpha_{2}+\alpha_{3} \right)^{2}\right]
            \\
            _{t}H^{\varepsilon}_{3} &
            \frac{1}{2}{\frac {{\epsilon}^{2} \left( {{1- \alpha_{2}}}^{2}\right) }{{ \alpha_{3}}{ \alpha_{2}}}} & 
            \frac{1}{2}{\frac {{\epsilon}^{2} \left( {{1- \alpha_{2}}}^{2}\right) }{{ \alpha_{3}}{ \alpha_{2}}}} &
            { \alpha_{2}} & 0 &  -1 & 0& 0& 0& 0 & 0 &
            \delta^{2}\alpha_{3}\left(1-\alpha_{2}^{2}\right)
            \\
            _{1}D_{2} & \frac{1}{2}{ \delta_{1}} & \frac{1}{2}{ \delta_{1}} & \frac{1}{2} & 
            -\frac{1}{2} & 0 & 0 & \frac{1}{2}-\frac{1}{4}({ \delta_{1}}-{ \delta_{2}}) & 
            \frac{1}{4}( \delta_{2}-\delta_{1})&
            -\frac{1}{4}(\delta_{1}+\delta_{2}) &
            \frac{1}{2}-\frac{1}{4}(\delta_{1}+\delta_{2})&
            0
            \\
            _{2}D_{2} & \frac{1}{2}& -\frac{1}{2} & \frac{1}{2}\delta_{1} & \frac{1}{2}\delta_{1} & 0 & 0 & 
            \frac{1}{2}-\frac{1}{4}( \delta_{1}- \delta_{2}+\delta_{1}\lambda) & 
            \frac{1}{4}(\delta_{1}\lambda-\delta_{1}+\delta_{2}) & 
            \frac{1}{2}-\frac{1}{4}(\delta_{1}-\delta_{1}\lambda+\delta_{2})&
            -\frac{1}{4}(\delta_{1}+\delta_{1}\lambda+\delta_{2})& 
            -\delta_{1} \delta_{2}\lambda
            \\
            _{3}D_{2} & \frac{1}{2} & -\frac{1}{2} & 0 & 0& \frac{1}{2}{ \delta_{1}} & 
            -\frac{1}{2}{ \delta_{1}} & 
            \frac{1}{2}-\frac{1}{4}(\delta_{1}-\delta_{2}+\delta_{1}\lambda) & 
            \frac{1}{4}( \delta_{1}+\delta_{1}\lambda+ \delta_{2}) & 
            \frac{1}{2}-\frac{1}{4}( \delta_{1}-\delta_{1}\lambda+ \delta_{2}) & 
            \frac{1}{4}(\delta_{1} -\delta_{1}\lambda-\delta_{2}) & 
            -{ \delta_{1}}{ \delta_{2}}\lambda
            \\
            D_{3} & \frac{1}{2} & \frac{1}{2} & \frac{1}{2} & \frac{1}{2} & \frac{1}{2} & -\frac{1}{2} & \frac{1}{4} & \frac{1}{4}&-\frac{1}{4} & 
            -\frac{1}{4} & 0
            \\
            _{1}D_{4} & \frac{1}{2}{ \delta_{2}}& \frac{1}{2}{ \delta_{2}} & 1 & 0 & \frac{1}{2}{ \delta_{1}} &
            -\frac{1}{2}{ \delta_{1}} & 0 & 0 & 0 & 0 & { \delta_{3}}
            \\
            _{2}D_{4} & 1 & 0 & \frac{1}{2}{ \delta_{2}} & \frac{1}{2}{ \delta_{2}} & \frac{1}{2}{ \delta_{1}} & 
            -\frac{1}{2}{ \delta_{1}}& 0& 0& 0& 0 & { \delta_{3}}
            \\
            \bottomrule
        \end{array}}
    \]
    \caption{Identification of the coefficients of the non autonomous $\QV$ equation
        with those of the Boll's equations (\ref{eq:H4rhombic},\ref{eq:trapezoidalH4},
        \ref{eq:h6}). Since $a_{1}=a_{2,i}=0$ for every equation these coefficients
    absent in the Table.}
    \label{tab:identification}
\end{sidewaystable}

\clearpage

\section{Conclusions}

In this article we propose a non autonomous extension of the $\QV$ equation (\ref{eq:naqv}) which satisfies an extended Klein symmetry (\ref{KL}) in such a way that in the autonomous sub-case reduces to the $\QV$ equation (\ref{eq:qv}). The so obtained non autonomous $\QV$ equation includes all equations of the Boll classification as its sub-cases and results integrable by the Algebraic Entropy test. Using the construction proposed by Xenitidis \cite{Xenitidis2009}  one builds up a symmetry for the QV non autonomous with respect to one direction only.  This symmetry turns out to belong to the class of the
 non autonomous YdKN equation proposed by Levi and Yamilov \cite{LeviYamilov1995}, an equation satisfying all five integrability conditions obtained by the formal generalized symmetry method. One is confident that the results obtained using the formulae (\ref{eq:ttxsymmn}, \ref{eq:ttxsymmm}) by Xenitidis \cite{Xenitidis2009} are correct also in general setting, as for all the sub-cases of the non autonomous QV belonging to the Boll classification the three point generalized symmetries, calculated using both the definition or through the formula by Xenitidis, coincide and belong to the class of the non autonomous YdKN. However we have not been able to have a complete direct proof in the general non autonomous case due to computational complexities and work on it is in progress. In all generality we have two possible connection formulae (\ref{c1}, \ref{c2}) between the 16 coefficients of the general non autonomous QV and the 9 coefficients of the non autonomous YdKN, one corresponding to the Xenitidis formula (\ref{eq:ttxsymmn}) along the $n$ direction and another one corresponding to (\ref{eq:ttxsymmm}) along $m$.
 
\section*{Acknowledgment}

\indent CS and DL  have been partly supported by the Italian Ministry of Education and Research, 2010 PRIN {\it Continuous and discrete nonlinear integrable evolutions: from water waves to symplectic maps}.

\noindent GG and DL are supported   by INFN   IS-CSN4 {\it Mathematical Methods of Nonlinear Physics}.

\appendix

\section{Explicit form the \Hvier~and \Hsechs~equations}

\label{app:equations}

Throughout this appendix we will use the notation:
\begin{equation}
    F^{\pm}_{k} = \frac{1\pm(-1)^{k}}{2}, \quad k \in \Z.
    \label{eq:fpmk}
\end{equation}

\paragraph{Rhombic \Hvier~equations:}
\begin{subequations}
    \begin{align}
        _{r}H_{1}^\epsilon &\colon
        \begin{aligned}[t]
            &\phantom{+}(u_{n,m}-u_{n+1,m+1})\, (u_{n+1,m}-u_{n,m+1})\, -\,(\alpha \,- \, \beta)\\
        &+\epsilon (\alpha - \beta)\left(\Fp{n+m} \,u_{n+1,m} u_{n,m+1} 
        + \Fm{n+m} \,u_{n,m} u_{n+1,m+1}\right)= 0,
        \end{aligned}
        \label{eq:H1ae} 
        \\
        _{r}H_2^\epsilon &\colon
        \begin{aligned}[t]
            &\phantom{+}(u_{n,m}-u_{n+1,m+1})(u_{n+1,m}-u_{n,m+1}) + \\
        &+(\beta-\alpha) (u_{n,m}+u_{n+1,m}+u_{n,m+1}+u_{n+1,m+1}) 
        - \alpha^2 + \beta^2  \\
        &-  \epsilon\, (\beta-\alpha)^3  -\epsilon \,(\beta-\alpha) \,\left( 2 \Fm{n+m} u_{n,m} 
        + 2 \Fp{n+m} u_{n+1,m}+\alpha+\beta\right) \cdot \\
        &\cdot \left(  2 \Fm{n+m} u_{n+1,m+1} + 2 \Fp{n+m} u_{n,m+1} +\alpha+\beta\right )\,=\,0, 
        \end{aligned}
        \label{eq:H2ae}
        \\
        _{r}H_3^\epsilon &\colon
        \begin{aligned}[t]
            &\phantom{+}\alpha (u_{n,m} u_{n+1,m}+u_{n,m+1} u_{n+1,m+1}) \\
        &- \beta (u_{n,m} u_{n,m+1}+u_{n+1,m} u_{n+1,m+1}) + (\alpha^2-\beta^2) \delta  \\
        &-\, \frac{\epsilon (\alpha^2-\beta^2)}{\alpha \beta}
        \left(\Fp{n+m} \,u_{n+1,m} u_{n,m+1} + \Fm{n+m} \,u_{n,m} u_{n+1,m+1}\right) = 0,
        \end{aligned}
        \label{eq:H3ae}
    \end{align}
    \label{eq:H4rhombic}
\end{subequations}

\paragraph{Trapezoidal \Hvier~equations:}
\begin{subequations}
    \begin{align}
        _{t}H_{1}\colon &
        \begin{aligned}[t]
        &\left(u_{n,m}-u_{n+1,m}\right)  \left(u_{n,m+1}-u_{n+1,m+1}\right)-\\ 
        &-\alpha_{2}\epsilon^2\left({F}_{m}^{\left(+\right)}u_{n,m+1}u_{n+1,m+1}
        +{F}_{m}^{\left(-\right)}u_{n,m}u_{n+1,m}\right) -\alpha_{2}=0,
        \end{aligned}
        \label{eq:tH1e}
        \\
        _{t}H_{2}\colon &
        \begin{aligned}[t]
        &\left(u_{n,m}-u_{n+1,m}\right)\left(u_{n,m+1}-u_{n+1,m+1}\right)
        \\
        &+\alpha_{2}\left(u_{n,m}+u_{n+1,m}+u_{n,m+1}+u_{n+1,m+1}\right)
        \\
        &+\frac{\epsilon\alpha_{2}}{2} \left(2{F}_{m}^{\left(+\right)}u_{n,m+1}
        +2\alpha_{3}+\alpha_{2}\right)\left(2{F}_{m}^{\left(+\right)}u_{n+1,m+1}+2\alpha_{3}+\alpha_{2}\right)
        \\
        &+\frac{\epsilon\alpha_{2}}{2} \left(2{F}_{m}^{\left(-\right)}u_{n,m}+2\alpha_{3}
        +\alpha_{2}\right)\left(2{F}_{m}^{\left(-\right)}u_{n+1,m}+2\alpha_{3}+\alpha_{2}\right)
        \\
        &+\left(\alpha_{3}+\alpha_{2}\right)^2
        -\alpha_{3}^2-2\epsilon\alpha_{2}\alpha_{3}\left(\alpha_{3}+\alpha_{2}\right)=0
        \end{aligned}
        \label{eq:tH2e}
        \\
        _{t}H_{3}\colon &
        \begin{aligned}[t]
        &\alpha_{2}\left(u_{n,m}u_{n+1,m+1}+u_{n+1,m}u_{n,m+1}\right)
        \\
        &-\left(u_{n,m}u_{n,m+1}+u_{n+1,m}u_{n+1,m+1}\right)
        -\alpha_{3}\left(\alpha_{2}^{2}-1\right)\delta^2+
        \\
        &-\frac{\epsilon^2(\alpha_{2}^{2}-1)}{\alpha_{3}\alpha_{2}}
        \left({{F}_{m}^{\left(+\right)}u_{n,m+1}u_{n+1,m+1}
        +{F}_{m}^{\left(-\right)}u_{n,m}u_{n+1,m}}\right)=0,
        \end{aligned}
        \label{eq:tH3e}
    \end{align}
    \label{eq:trapezoidalH4}
\end{subequations}

\paragraph{\Hsechs~equations:}
\begin{subequations}
    \begin{align}
        _{1}D_{2} &\colon
        \begin{aligned}[t]
        &\phantom{+}\left( F_{n+m}^{\left(-\right)}-\delta_{1} F_{n}^{\left(+\right)} F_{m}^{\left(-\right)}+\delta_{2} F_{n}^{\left(+\right)} F_{m}^{\left(+\right)}\right)u_{n,m}
        \\
        &+\left( F_{n+m}^{\left(+\right)}-\delta_{1} F_{n}^{\left(-\right)} F_{m}^{\left(-\right)}+\delta_{2} F_{n}^{\left(-\right)} F_{m}^{\left(+\right)}\right)u_{n+1,m}+
        \\ 
        &+\left( F_{n+m}^{\left(+\right)}-\delta_{1} F_{n}^{\left(+\right)} F_{m}^{\left(+\right)}+\delta_{2} F_{n}^{\left(+\right)} F_{m}^{\left(-\right)}\right)u_{n,m+1}
        \\
        &+\left( F_{n+m}^{\left(-\right)}-\delta_{1} F_{n}^{\left(-\right)} F_{m}^{\left(+\right)}+\delta_{2} F_{n}^{\left(-\right)} F_{m}^{\left(-\right)}\right)u_{n+1,m+1}+
        \\ 
        &+\delta_{1}\left( F_{m}^{\left(-\right)}u_{n,m}u_{n+1,m}+ F_{m}^{\left(+\right)}u_{n,m+1}u_{n+1,m+1}\right)
        \\
        &+ F_{n+m}^{\left(+\right)}u_{n,m}u_{n+1,m+1}
        + F_{n+m}^{\left(-\right)}u_{n+1,m}u_{n,m+1}=0,
        \end{aligned}
        \label{eq:1D2}
        \\
        _{2}D_{2} &\colon
        \begin{aligned}[t]
            &\phantom{+}\left(F_{m}^{\left(-\right)}-\delta_{1}F_{n}^{\left(+\right)}F_{m}^{\left(-\right)}+\delta_{2}F_{n}^{\left(+\right)}F_{m}^{\left(+\right)}-\delta_{1} \lambda F_{n}^{\left(-\right)}F_{m}^{\left(+\right)}\right)u_{n,m}
        \\
        &+\left(F_{m}^{\left(-\right)}-\delta_{1}F_{n}^{\left(-\right)}F_{m}^{\left(-\right)}+\delta_{2}F_{n}^{\left(-\right)}F_{m}^{\left(+\right)}-\delta_{1} \lambda F_{n}^{\left(+\right)}F_{m}^{\left(+\right)}\right)u_{n+1,m}
        \\
        &+\left(F_{m}^{\left(+\right)}-\delta_{1}F_{n}^{\left(+\right)}F_{m}^{\left(+\right)}+\delta_{2}F_{n}^{\left(+\right)}F_{m}^{\left(-\right)}-\delta_{1} \lambda F_{n}^{\left(-\right)}F_{m}^{\left(-\right)}\right)u_{n,m+1}
        \\
        &+\left(F_{m}^{\left(+\right)}-\delta_{1}F_{n}^{\left(-\right)}F_{m}^{\left(+\right)}+\delta_{2}F_{n}^{\left(-\right)}F_{m}^{\left(-\right)}-\delta_{1} \lambda F_{n}^{\left(+\right)}F_{m}^{\left(-\right)}\right)u_{n+1,m+1}
        \\
        &+\delta_{1}\left(F_{n+m}^{\left(-\right)}u_{n,m}u_{n+1,m+1}+F_{n+m}^{\left(+\right)}u_{n+1,m}u_{n,m+1}\right)
        \\ 
        &+F_{m}^{\left(+\right)}u_{n,m}u_{n+1,m}+F_{m}^{\left(-\right)}u_{n,m+1}u_{n+1,m+1}
        -\delta_{1}\delta_{2}\lambda=0,
        \end{aligned}
        \label{eq:2D2}
        \\
        _{3}D_{2} &\colon
        \begin{aligned}[t]
            &\phantom{+}\left(F_{m}^{\left(-\right)}-\delta_{1}F_{n}^{\left(-\right)}F_{m}^{\left(-\right)}+\delta_{2}F_{n}^{\left(+\right)}F_{m}^{\left(+\right)}-\delta_{1} \lambda F_{n}^{\left(-\right)}F_{m}^{\left(+\right)}\right)u_{n,m}
        \\
        &+\left(F_{m}^{\left(-\right)}-\delta_{1}F_{n}^{\left(+\right)}F_{m}^{\left(-\right)}+\delta_{2}F_{n}^{\left(-\right)}F_{m}^{\left(+\right)}-\delta_{1} \lambda F_{n}^{\left(+\right)}F_{m}^{\left(+\right)}\right)u_{n+1,m}
        \\
        &+\left(F_{m}^{\left(+\right)}-\delta_{1}F_{n}^{\left(-\right)}F_{m}^{\left(+\right)}+\delta_{2}F_{n}^{\left(+\right)}F_{m}^{\left(-\right)}-\delta_{1} \lambda F_{n}^{\left(-\right)}F_{m}^{\left(-\right)}\right)u_{n,m+1}
        \\
        &+\left(F_{m}^{\left(+\right)}-\delta_{1}F_{n}^{\left(+\right)}F_{m}^{\left(+\right)}+\delta_{2}F_{n}^{\left(-\right)}F_{m}^{\left(-\right)}-\delta_{1} \lambda F_{n}^{\left(+\right)}F_{m}^{\left(-\right)}\right)u_{n+1,m+1}
        \\
        &+\delta_{1}\left(F_{n}^{\left(-\right)}u_{n,m}u_{n,m+1}+F_{n}^{\left(+\right)}u_{n+1,m}u_{n+1,m+1}\right) 
        \\ 
        &+F_{m}^{\left(-\right)}u_{n,m+1}u_{n+1,m+1}
        +F_{m}^{\left(+\right)}u_{n,m}u_{n+1,m}-\delta_{1}\delta_{2}\lambda=0,
        \end{aligned}
        \label{eq:3D2}
        \\
        D_{3} &\colon
        \begin{aligned}[t]
            &\phantom{+}F_{n}^{\left(+\right)}F_{m}^{\left(+\right)}u_{n,m}+F_{n}^{\left(-\right)}F_{m}^{\left(+\right)}u_{n+1,m}
            +F_{n}^{\left(+\right)}F_{m}^{\left(-\right)}u_{n,m+1}
            \\
            &+F_{n}^{\left(-\right)}F_{m}^{\left(-\right)}u_{n+1,m+1}
            +F_{m}^{\left(-\right)}u_{n,m}u_{n+1,m}
            \\
            &+F_{n}^{\left(-\right)}u_{n,m}u_{n,m+1}+F_{n+m}^{\left(-\right)}u_{n,m}u_{n+1,m+1}+
        \\
        &+F_{n+m}^{\left(+\right)}u_{n+1,m}u_{n,m+1}+F_{n}^{\left(+\right)}u_{n+1,m}u_{n+1,m+1}
        \\
        &+F_{m}^{\left(+\right)}u_{n,m+1}u_{n+1,m+1}=0,
        \end{aligned}
        \label{eq:D3}
        \\
        _{1}D_{4} &\colon
        \begin{aligned}[t]
            &\phantom{+}\delta_{1}\left(F_{n}^{\left(-\right)}u_{n,m}u_{n,m+1}+F_{n}^{\left(+\right)}u_{n+1,m}u_{n+1,m+1}\right)+\\
            &+\delta_{2}\left(F_{m}^{\left(-\right)}u_{n,m}u_{n+1,m}+F_{m}^{\left(+\right)}u_{n,m+1}u_{n+1,m+1}\right)+\\
            &+u_{n,m}u_{n+1,m+1}+u_{n+1,m}u_{n,m+1}+\delta_{3}=0,
        \end{aligned}
        \label{eq:1D4}
        \\
        _{2}D_{4} &\colon
        \begin{aligned}[t]
            &\phantom{+}\delta_{1}\left(F_{n}^{\left(-\right)}u_{n,m}u_{n,m+1}+F_{n}^{\left(+\right)}u_{n+1,m}u_{n+1,m+1}\right)+
            \\
            &+\delta_{2}\left(F_{n+m}^{\left(-\right)}u_{n,m}u_{n+1,m+1}+F_{n+m}^{\left(+\right)}u_{n+1,m}u_{n,m+1}\right)+
            \\
            &+u_{n,m}u_{n+1,m}+u_{n,m+1}u_{n+1,m+1}+\delta_{3}=0.
        \end{aligned}
        \label{eq:2D4}
    \end{align}
    \label{eq:h6}
\end{subequations}

\section{Correlation formulae between the non autonomous $\QV$ and the non autonomous YdKN in the $m$ direction.}

\label{app:QV}
For the sake of completeness let us write down  the non autonomous YdKN in the $m$ direction:
\begin{equation}
    \diff{u_{m}}{t} = \frac{A_{m}(u_{m})u_{m+1}u_{m-1}
    + B_{m}(u_{m})(u_{m+1}+u_{m-1}) + C_{m}(u_{m})}{u_{m+1}-u_{m-1}}.
    \label{eq:naydkm}
\end{equation}
Here the $m$-dependent coefficients are given by:
\begin{subequations}
\begin{align}
    A_{m}(u_m)&= a u_{m}^{2} + 2 b_{m} u_{m} + c_{m},
    \\
    B_{m}(u_m) &= b_{m+1} u_{m}^{2} + d u_{m} + e_{m+1},
    \\
    C_{m} (u_m)&= c_{m+1}u_{m}^{2} + 2 e_{m} u_{m} + f,
\end{align}
    \label{eq:akbkcknaydkm}%
\end{subequations}
where $b_{m}$, $c_{m}$ and $e_{m}$ are 2-periodic functions, i.e. 
\bea \label{aYdKNm}
b_m=b_0+b_1 (-1)^m,\; c_m=c_0+c_1 (-1)^m, \; e_m=e_0+e_1 (-1)^m. 
\eea 
The correlation formulae read:
\bea \label{c2}
a&=&a_1 a_{5,0}- a_{2,0}^2  + \\ \nonumber && \quad -a_{2,1}^2 +  a_{2,2}^2 + 
  a_{2,3}^2 + (-1)^n (2 a_{2,0} a_{2,1} - 2  a_{2,2} a_{2,3}  +   a_1 a_{5,1}), \\ \nonumber b_0&=&\frac 1 2 \{a_{2,0} (a_{5,0} -  a_{3,0} -  a_{4,0}) + 
 a_1 a_{6,0} - a_{2,2} a_{3,2}   - a_{2,3} a_{4,3} +a_{2,1} a_{5,1} + \\ \nonumber &&\; + (-1)^n[ a_{2,1} (a_{5,0}  +  a_{3,0}+ a_{4,0})+a_{2,3} a_{3,2}  + a_1 a_{6,1}  + a_{2,0} a_{5,1}  +\\ \nonumber && \quad + a_{2,2} a_{4,3}]\}, \\ \nonumber 
 b_1&=&\frac 1 2 \{ a_{2,2} (a_{4,0}  -  a_{3,0}-  a_{5,0}) - a_{2,3} a_{5,1} + a_{2,1} a_{4,3}   - a_{2,0} a_{3,2}  + 
 a_1 a_{6,2} +\\ \nonumber && \; + (-1)^n [a_1 a_{6,3} +  a_{2,3} (a_{3,0} - a_{4,0}-  a_{5,0}) + a_{2,1} a_{3,2} -  a_{2,0} a_{4,3}-\\ \nonumber && \quad -  a_{2,2} a_{5,1}] \},\\ \nonumber 
 c_0&=&a_{2,0} a_{6,0} -a_{4,0} a_{3,0} 
+ a_{2,1} a_{6,1} - a_{2,3} a_{6,3}- 
 a_{2,2} a_{6,2}+\\ \nonumber &&\; - (-1)^n [a_{2,2} a_{6,3} +  a_{4,3} a_{3,2}  + a_{2,3} a_{6,2} - a_{2,0} a_{6,1}  -  a_{2,1} a_{6,0} ], \\ \nonumber
 c_1&=& a_{2,1} a_{6,3}-a_{4,0} a_{3,2}+   
 a_{2,0} a_{6,2}-a_{2,3} a_{6,1}-a_{2,2} a_{6,0}-\\ \nonumber && \; -(-1)^n[a_{2,2} a_{6,1}+ a_{4,3} a_{3,0}   + a_{2,3} a_{6,0}  -   a_{2,1} a_{6,2}  
  - a_{2,0} a_{6,3}], \\ \nonumber 
  d&=&\frac 1 2 [a_{5,0}^2 - a_{4,0}^2- a_{3,0}^2+ a_1 a_7+ a_{3,2}^2  + a_{4,3}^2  - a_{5,1}^2 +\\ \nonumber && \; + 4 (-1)^n (a_{2,1} a_{6,0} + a_{2,0} a_{6,1} + a_{2,3} a_{6,2} + a_{2,2} a_{6,3}) ], \\ \nonumber 
  e_0&=&\frac 1 2 \{ a_{6,0} [a_{5,0}- a_{4,0}  - a_{3,0} ]+ a_{2,0} a_7- 
 a_{5,1} a_{6,1}+
 a_{3,2} a_{6,2}+ 
 a_{4,3} a_{6,3}-\\ \nonumber && \; - (-1)^n [a_{3,2} a_{6,3}  +  a_{4,3} a_{6,2} +  a_{5,1} a_{6,0} - a_{6,1} (a_{3,0}   + a_{4,0}  + a_{5,0} )    -  a_{2,1} a_7]\},\\ \nonumber 
 e_1&=&\frac 1 2 \{ -a_{6,2}[a_{3,0}- a_{4,0}+ 
 a_{5,0} ] + a_{3,2} a_{6,0}- a_{4,3} a_{6,1}+
 a_{5,1} a_{6,3}+a_{2,2} a_7+\\ \nonumber && \;+(-1)^n[ a_{4,3} a_{6,0}   -  a_{3,2} a_{6,1}   +  a_{5,1} a_{6,2}   +   a_{6,3}(a_{3,0} -  a_{5,0}-  a_{4,0})  +  a_{2,3} a_7\}, \\ \nonumber 
 f&=& 
 a_{5,0} a_7-a_{6,0}^2+ a_{6,2}^2+ a_{6,3}^2- a_{6,1}^2-(-1)^n(2  a_{6,2} a_{6,3}-  2  a_{6,0} a_{6,1}  +  a_{5,1} a_7).
  \eea

\end{document}